**Noise-dependent bias in quantitative STEM-EMCD experiments revealed by bootstrapping**

Hasan Ali[1,2,5,*], Jan Rusz[3], Daniel E. Bürgler[4], Roman Adam[4], Claus M. Schneider[4], Cheuk-Wai Tai[2], Thomas Thersleff[2]

[1] Department of Materials Science and Engineering, Uppsala University, Box 534, 751 21, Uppsala, Sweden

[2] Department of Materials and Environmental Chemistry, Stockholm University, 106 91 Stockholm, Sweden

[3] Department of Physics and Astronomy, Uppsala University, Box 516, 751 20 Uppsala, Sweden

[4] Peter Grünberg Institut, D-52425, Forschungszentrum Jülich GmbH, Jülich, Germany

[5] Ernst Ruska-Centre for Microscopy and Spectroscopy with Electrons, Forschungszentrum Jülich, 52425 Jülich, Germany

[*]Corresponding author: ali.hasan@angstrom.uu.se

**Abstract**

Electron magnetic circular dichroism (EMCD) is a powerful technique for estimating element-specific magnetic moments of materials on nanoscale with the potential to reach atomic resolution in transmission electron microscopes. However, the fundamentally weak EMCD signal strength complicates quantification of magnetic moments, as this requires very high precision, especially in the denominator of the sum rules. Here, we employ a statistical resampling technique known as bootstrapping to an experimental EMCD dataset to produce an empirical estimate of the noise-dependent error distribution resulting from application of EMCD sum rules to bcc iron in a 3-beam orientation. We observe clear experimental evidence that noisy EMCD signals preferentially bias the estimation of magnetic moments, further supporting this with error distributions produced by Monte-Carlo simulations. Finally, we propose guidelines for the recognition and minimization of this bias in the estimation of magnetic moments.

## 1. Introduction

Electron magnetic circular dichroism (EMCD) [1] is a transmission electron microscopy (TEM) technique that utilizes electron energy loss spectroscopy (EELS) to measure the element-specific magnetic moments of ferro- or ferrimagnetic materials. EMCD is the electron equivalent of the well-established X-ray based technique x-ray magnetic circular dichroism XMCD [2], with the key difference that the conjugated momentum transfer relies on the crystal splitting in EMCD rather than a circularly-polarized radiation source. While EMCD can measure the element specific magnetic moments of the materials, it also offers the highest possible spatial resolution [3] [4] [5] [6] with sufficient depth of analysis, something which is not possible with XMCD and other techniques due to either limited spatial resolution [7] [8] or insufficient depth resolution [9] [10]. Recently, differential phase contrast (DPC) microscopy has achieved atomic resolution magnetic measurements [11] but this requires a TEM with a custom-designed objective lens system [12] to obtain atomic resolution electron probes at the specimen under magnetic field free environment. Additionally, DPC lacks the ability to quantify spin and orbital magnetic moments, while these critical properties can be extracted from EMCD experiments through application of EMCD sum rules [13]. A comprehensive review on EMCD can be found here [14].

Although EMCD comes with many attractive advantages over the complementary techniques for magnetic characterization, its hallmark feature – the quantitative estimation of orbital and spin magnetic moments – is particularly challenging. As initially described by Schattschneider et al. [1], an EMCD signal is experimentally obtained by taking the difference of EELS spectra acquired at conjugated scattering angles across the diffracted spots in the reciprocal space. In practice, these scattering angles lie at off-axis positions far away from the Bragg spots, yielding a signal that can be orders of magnitude smaller than standard EELS experiments, thereby resulting in a notoriously low signal to noise ratio (SNR). While this design is sufficient to reveal modifications to the selection rules governing 2p to 3d transitions in transition metals caused by the presence of an uncompensated magnetic field, a quantitative estimation of magnetic moments requires application of the EMCD sum rules [13] [15]. This involves significant signal processing steps including deconvolution [16], continuum removal (typically achieved by subtracting two conjugated signals under the assumption of ideal symmetrical scattering conditions), integration, and numerical manipulation, with each additional processing step introducing additional errors to the already noisy signal. Since the ratio of magnetic spin to orbital moments ($m_L/m_S$) is very small for most of the technologically important magnetic materials, minor errors in the quantification process risk interpretation as novel magnetic properties. Despite many efforts to improve the EMCD signal strength [16] [17] [18] [19] [20] poor SNR remains a major challenge for wider adoption of the EMCD method.

Although the influence and analysis of statistical errors is extremely important in EMCD experiments, most published studies reporting error bars have derived them independently using individualized approaches [3] [5] [21] [22] [23] [24]. These studies either estimate statistical errors from a very small sample size [5], or they analytically derive them for single EMCD spectra, typically using residuals in the pre-edge or post-edge region of the EELS spectra as estimators for the statistical variance, and then propagating them appropriately [6] [20] [25] [26] [27] [28]. While this may be appropriate given the small number of studies and individual spectra available to scientists at the time of publication, it can also be viewed as problematic from the perspective of the wider community for the following reasons. These approaches all implicitly assume that the errors follow a normal distribution cantered on the calculated value, and that this distribution function does not change as a function of the initial noise corruption of the original data. These assumptions have not yet been substantiated with supportive evidence and, thus, should be publicly challenged, especially considering the complicated procedure required to apply the sum rules. Moreover, the errors determined in many EMCD papers [5] [6] [23] [24] [26] [29] [30] are comparable to XMCD experiments [31] [32] despite the latter technique boasting a significantly higher SNR.

One way to challenge the validity of the assumptions used to justify the analytical, theory-centric derivation of statistical errors outlined in the papers above is to take an empirical experiment-centric approach. In principle, such an empirical approach could be achieved by independently measuring $m_L/m_S$ many thousands of times from the exact same region of material under a range of different SNR conditions. Multiple measurements at a given SNR could then be fit to a distribution function to provide a more complete description of statistical errors (also subverting the need to worry about systematic errors), while different SNR values could reveal any noise-dependent variations to this distribution function. However, in practice, this experimental design is greatly complicated by the highly localized nature of EMCD itself as well as the high doses inherent in such experiments. Extracting the EMCD signal from exactly the same region of the sample for the time necessary for this experiment would inevitably lead to time-dependent beam damage problems in even the most radiation-robust samples. Also, if a large, parallel probe is used, exposing fresh sample regions for each measurement introduces the prospect of systematic errors arising from variations due to crystal tilt [25] and sample thickness [33], in addition to placing unreasonably large burden on the homogeneity of the sample preparation and growth. Thus, the experimental design needed to achieve the ideal conditions to take the empirical approach is time and cost prohibitive.

In this paper, we argue that, by combining recent advances in EMCD experimental design with a well-established statistical inference procedure broadly known as bootstrapping, we can approximate the conditions needed for an empirical error analysis reasonably closely, thereby allowing us to study and discuss the aptness of the aforementioned assumptions. The experimental design we employ is known as STEM-EMCD and has been described previously [3], while bootstrapping is a resampling-with-replacement technique commonly used to estimate the statistical error distribution from sample-limited datasets [34]. Although bootstrapping has previously been applied to various electron microscopy experiments [35] [36] [37], its use in this experiment specifically allows us to test the dependence of the error distribution shape on the SNR of EMCD signal, which has not yet been investigated. Critically, we observe that one of the main assumptions justifying the analytical approach to error analysis – that of symmetric, normally-distributed errors – is not appropriate for lower SNR values. This leads to an obvious noise-dependent bias towards higher values of $m_L/m_S$ that can have major ramifications for the most high-impact studies, where signal is at an absolute premium and high SNR may simply be infeasible. Moreover, from the Monte-Carlo simulations, we find that this noise dependent bias is also material dependent and for the same noise levels, materials having larger values of $m_L/m_S$ show higher bias. Importantly, the approach we outline here provides us with sufficient information to mathematically describe and account for this bias, permitting us to propose cautionary guidelines for the interpretation of future quantitative EMCD results. The methods and workflow proposed here are also transferrable to other magnetic materials and can be generalized to other STEM-based experiments or even XMCD error analysis under the right conditions.

## 2. Methods

### 2.1. Sample Fabrication

The sample used in this experiment was fabricated in the same way as in [38]. The fabrication was carried out by thermally evaporating a 40 nm thick bcc-Fe layer and a 3 nm thick Al top layer onto a 5 nm thick $Si_3N_4$ membrane under ultra-high vacuum conditions. The membrane was used without prior cleaning and kept at room temperature during deposition. The layer thicknesses were monitored with calibrated quartz microbalances and are estimated to fluctuate by about 3%. To achieve larger lateral Fe grain sizes exceeding 100 nm, the Fe film was annealed immediately after deposition for 2h at 750°C. Subsequently, the Al cap layer was deposited at room temperature to protect the Fe from oxidation upon exposure to air.

### 2.2. STEM EMCD

To apply the bootstrapping technique on the EMCD analysis, the STEM-EMCD technique was employed [39] [40] [41] [42] using a convergence semi-angle of 8.0 mrad at 300 kV. First, a thin single-crystalline grain of iron was identified and tilted to the three-beam condition having the set of systematic row vectors $g = \langle 0\ 02 \rangle$. The grain itself was specifically chosen to be relatively featureless to facilitate a similar EMCD signal from each pixel position. Tilting was performed manually by inspecting the CBED pattern from a region adjacent to the ROI. The ROI as well as a corresponding CBED pattern are provided in **Fig 1(a)**.

Once the diffraction conditions were configured, the magnification of the projector system was adjusted to yield an EELS collection semiangle of 3.7 mrad at the 2.5 mm spectrometer entrance aperture. A new cartesian coordinate system was defined in which the centres of the (0 0 0) and (0 0 2) Bragg disks were defined as $x_0, y_0 = [0, 0]$ and $x_g, y_g = [1, 0]$ using a custom written script. Thus, $x$ describes the direction along the systematic row vector while $y$ describes the orthogonal direction, with the units for both normalized to 13.7 mrad (the scattering angle of Fe (0 0 2) lattice planes at 300 kV). Four chiral positions were set in this coordinate system following all sign permutations of the coordinates $x_{ij}, y_{ij} = [\pm 0.5, \pm 0.9]$. These positions are labelled according to their sign combination as shown in **Fig 1(a)**.

An ROI measuring 114.0 × 104.5 nm was defined with a step size of 0.38 nm (denoted in **Fig 1(b)**), yielding a 300 × 275 EELS spectrum image (ESI). Prior to each scan, the initial probe position was corrected using a survey image that was acquired with the HAADF detector at the outset of the experiment, thereby ensuring a reasonable spatial registration between scans at this resolution. The collection angles for each scan were set by using the diffraction shift deflectors to offset the diffraction pattern onto the spectrometer entrance aperture according to the $x, y$ convention for chiral positions established above. Following this, the ROI was scanned a total of 5 times in the order "++", "+-", "-+", "--", "00". The final scan ("00") was performed on-axis and was used to acquire the low-loss EELS dataset to allow for deconvolution as well as to estimate the region thickness, which is presented in **Fig 1(c)**. The spectrometer was configured using a dispersion of 0.25 eV / channel and an offset of 400 eV, allowing for both the iron and oxygen edges to be captured. During acquisition, the drift tube was excited using a sawtooth waveform to continuously shift the recording position of the EEL spectrum on the spectrometer scintillator, a technique known as gain averaging [43]. The dwell time for each spectrum was 5 ms and a high quality gain reference was acquired for each data-cube post-acquisition [44].

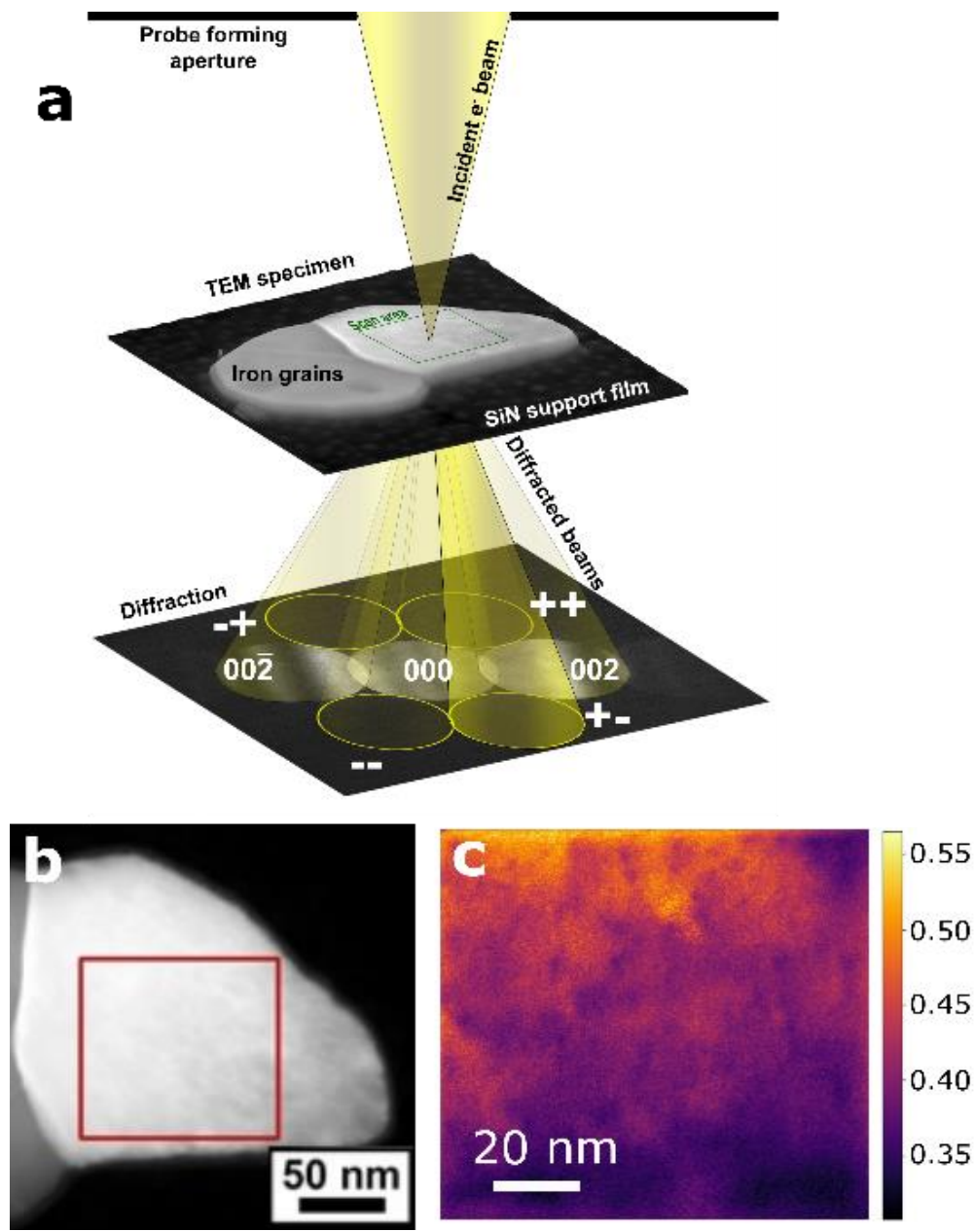

**Fig 1.** (a) A schematic showing the STEM-EMCD experimental setup, a fine electron probe is raster scanned across a well-defined region of the sample and an EELS spectrum is acquired at each probe position. The process is repeated four times, for the four aperture positions shown in the diffraction plane. The STEM-HAADF image and the diffraction pattern are taken from the real experimental data (b) HAADF image showing the area of the specimen scanned in the STEM-EMCD experiment (c) Thickness map for the scanned area extracted from low-loss EELS dataset. The colour bar shows the t/λ values.

### 2.3. Pre-processing of STEM-EMCD data

Pre-processing of the STEM-EMCD data was performed following the procedure outlined in Thersleff et al. [38]. First, strong pixel-level outliers were removed. Second, all datasets were concatenated along their y-axis direction. An EELS spectrum from the first dataset was chosen and used a reference to which all other spectra were aligned using cross-correlation. The shift correction roughly followed the sawtooth waveform used for the gain averaging and any outliers were corrected using this.

### 2.4. Bootstrapping workflow

STEM-EMCD is primarily advantageous in that it fractionates the beam dose over the scanning region. Integration of all recorded spectra within each chiral EELS spectrum image (ESI) datacube subsequently maximizes the SNR of the resulting EMCD while minimizing the influence of beam damage. However, it also means that each ESI can be subsampled to estimate lower SNR values. We take advantage of this property to study the SNR dependence of statistical EMCD errors here with bootstrapping.

For each ESI ($S_\pm$), we define the number of recorded spectra as $N_{px}$, representing the number of spatial "pixels." $N_{px}$ can also be understood as the vectorised 2D image, or $N_{px} = N_x \cdot N_y$. As discussed in the experimental section above, for this experiment, $N_{px} = 300 \cdot 275 = 82500$. We now define a subsampling parameter $N_s$, also called "specsum," which denotes the number of spectra that are integrated for each bootstrapping iteration. $N_s$ roughly corresponds to a SNR value, with the maximum SNR occurring when $N_s = N_{px}$. For each $N_s$ value, a random selection of $N_s$ of spatial indices is generated, which are subsequently integrated into their corresponding chiral spectra $f_\pm$. This integration is repeated $N$ times for each value of $N_s$, resulting in $N$ independent estimates of $f_\pm$ for that given value of $N_s$ (which, again, corresponds to an estimate of the SNR). If we define $I_n$ as the 2D array of summation indices in the range $n = 1, 2, \cdots, N$, then

$$f_{\pm,n} = \sum S_\pm I_n$$

Critically, since bootstrapping involves sampling with replacement, the same raw spectra may be chosen multiple times. This allows for a reliable estimate of error distributions even when $N_s = N_{px}$. This process is illustrated for clarity in **Fig 2**. Each column denotes a chosen value of $N_s$ (specsum), while each row shows a bootstrapping sample $n$ for that $N_s$. The maps depict $I_n$, which is a 2D array of the selected indices used for summation. As can be seen, the same index can be selected multiple times, which is graphically represented as a color denoted by the corresponding colorbar. The resulting spectrum from each $N_s, n$ combination reveals a corresponding EMCD signal, extracted using the methods described above. This is used to take a single estimate of $\frac{m_L}{m_S}$. For each $N_s$, a total of $N$ estimates are therefore obtained, and their distribution is plotted in the final row in the form of a 1D histogram. In the experiments we report here, $N = 1000$, and the resulting histogram is subsequently fit to a probability distribution function (PDF) to reliably estimate the distribution of errors in $\frac{m_L}{m_S}$ for that given SNR.

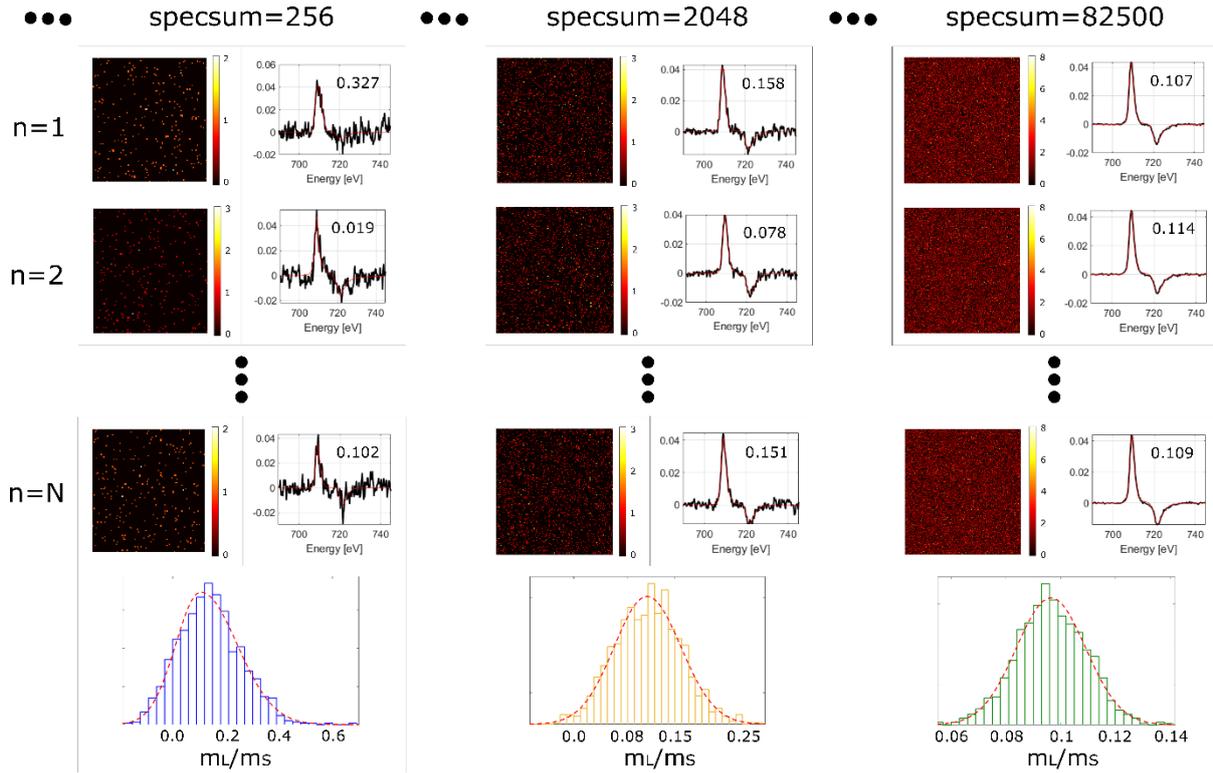

**Fig 2.** An illustration of bootstrapping workflow. For each value of $N_s$ (called specsum here), the spectra are randomly chosen and summed up from the chiral ESI datasets and the EMCD signal is obtained by processing and subtracting these spectra. The procedure is repeated for N=1000 times. The histograms show the distribution of the obtained 1000 $m_L/m_S$ values for each value of specsum.

## 3. Results

We produce various SNR conditions for EMCD analysis by integrating different number of raw EELS spectra (specsum). The application of bootstrapping workflow as described above on the experimental ESI datasets produces 1000 EMCD signals for each value of SNR (specsum) which is determined for both $L_3$ and $L_2$ energy loss edges of Fe. The EMCD signals are subsequently passed through the quantification workflow described in ref. [38] to obtain the corresponding $m_L/m_S$ values. Consequently, we have a dataset where for each SNR value, we have a distribution of 1000 $m_L/m_S$ values. The best way to visualize all these distributions as a function of SNR is a violin plot. In such a plot, each distribution is represented as a violin where the mean and median of each distribution are presented by a bar and a dot respectively. The height of each violin indicates the dispersion in each dataset as shown in **Fig 3**.

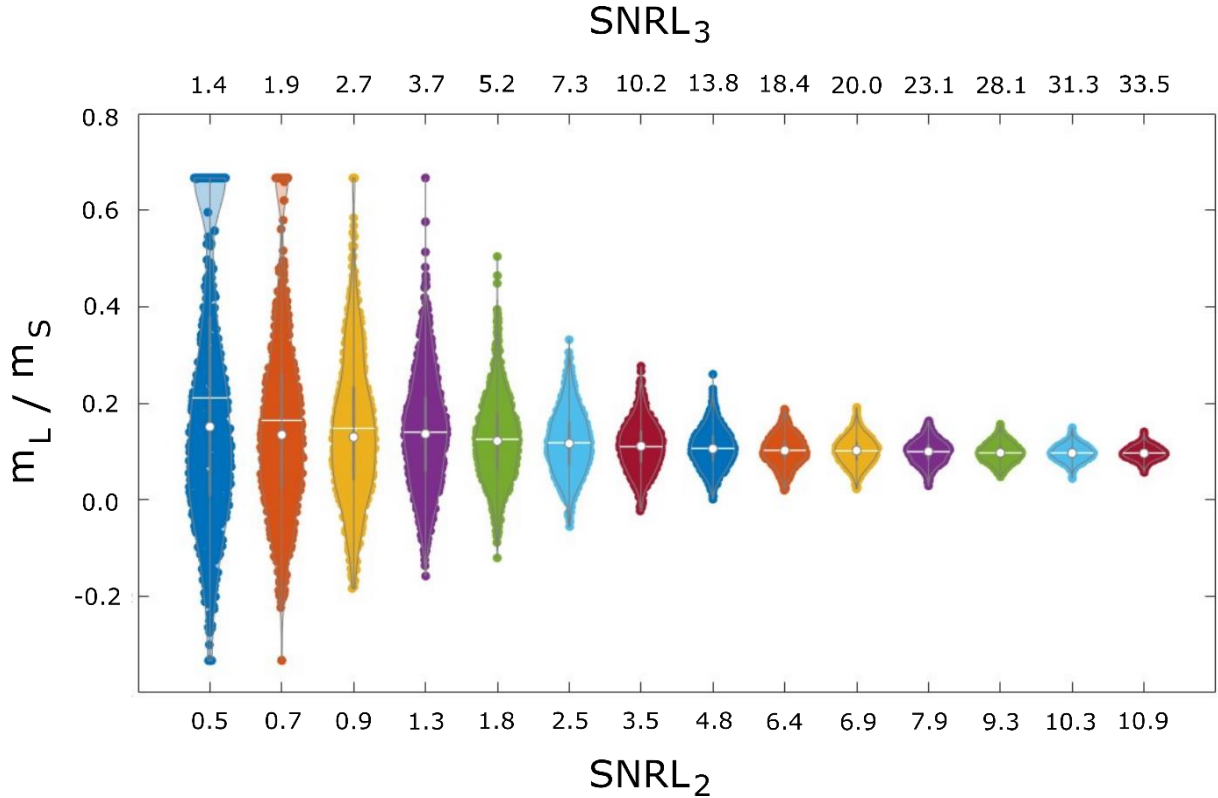

**Fig 3.** A violin plot showing the distribution of $m_L/m_S$ values as a function of signal to noise ratio at $L_3$ and $L_2$ energy loss edges. The white bar and dot in each violin indicate the mean and median of the distribution.

The first and obvious observation from **Fig 3** is that the dispersion in the resulting $m_L/m_S$ values gets higher as the SNR of the EMCD signal gets lower. The mean value of $m_L/m_S$ for each SNR is indicated by a white bar in each distribution. It is worth noting that the error bars which can be estimated by the height of each violin seems to be much higher than the typical error bars reported in EMCD literature and approach to ± 0.5 for EMCD signals with low SNRs, which is about 5 times the typical $m_L/m_S$ value for bcc Fe. This is huge and a single or few measurements with such a SNR can easily mislead the quantification results.

Another assumption made in the EMCD experiments is that the error bars are symmetrical and follow a Normal distribution around the mean value measured in any experiment. While fitting the $m_L/m_S$ distributions shown in **Fig 3**, we find that a simple Normal distribution function fails to fit the histograms as the SNR value decreases. By carefully fitting different probability distribution functions (PDF) to the results, we find that an epsilon skew-Normal distribution (ESD) [45] fits best to the results and it can be clearly seen in the left most fitted histogram in **Fig 2**. The ESD belongs to the family of asymmetric distribution functions consisting of a location ($\theta$), scale ($\sigma$) and skewness ($\varepsilon$) parameter, and it reduces to the Normal distribution when $\varepsilon = 0$. This noise dependent skewness is also reflected in the mean $m_L/m_S$ value for each distribution and biases the mean to larger values as shown in **Fig 3**. It means that the $m_L/m_S$ values would presumably tend to be higher than the real values for noisy EMCD signals.

Considering that the quantitative EMCD analysis is carried out by taking the ratio of the summation/differences seen at $L_3$ and $L_2$ edges (Equation 1), one of the reasons for this noise-dependent bias in the resulting $m_L/m_S$ values can be the asymmetric degradation of EMCD signal at $L_3$ and $L_2$ edges. The relatively lower signal seen at $L_2$ compared to $L_3$ edge causes a higher and quicker degradation of $SNR_{L_2}$ than $SNR_{L_3}$ for the same noise levels. It is also evident from **Fig 3** that the $SNR_{L_2}$ is much lower than $SNR_{L_3}$ for the same EMCD signals. Furthermore, the $L_2$ energy loss edge is usually broader than the L₃ edge, spreading over more channels on the CCD, accumulating more readout noise, resulting in faster deterioration of L₂ EMCD signal. It would be interesting to verify this claim by comparing the EMCD signals acquired on CCD and direct electron detectors. Nevertheless, looking at the violin plots shown in **Fig 3**, it is immediately clear that a simple median is a much better predictor of the "true" $m_L/m_S$ than the mean, and this gets pronounced for lower SNR values. To verify and support our experimental observations, we carried out Monte-Carlo simulations described below.

The error distribution for $m_L/m_S$ can be empirically determined using a Monte Carlo approach, which is described here. We begin by restating the sum rules [13] [46] :

$$\frac{m_L}{m_S} = \frac{2}{3} \frac{\int_{L_3} \Delta\sigma(E) dE + \int_{L_2} \Delta\sigma(E) dE}{\int_{L_3} \Delta\sigma(E) dE - 2\int_{L_2} \Delta\sigma(E) dE} \qquad (1)$$

As shown in Muto et al. [47], **Equation (1)** can be simplified by taking the full integral of the EMCD signal. This results in **Equation (2)** where $p$ and $q$ are defined in **Fig 4**.

$$\frac{m_L}{m_S} = \frac{2q}{9p - 6q} \qquad (2)$$

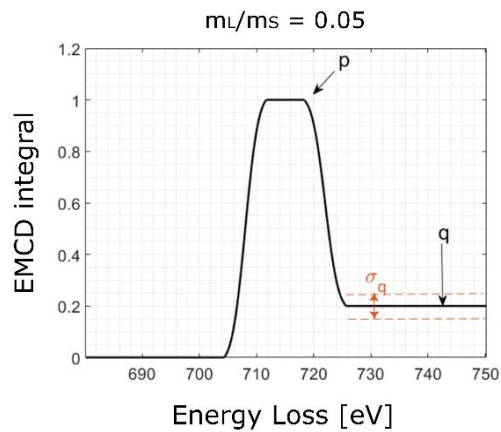

Fig 4. An integral of the EMCD signal showing the p and q values used in equation (2).

As in Muto et al. [47], we adopt **Equation (2)** and set $p = 1$, allowing $m_L/m_S$ to act as a function of $q$ alone. We now assume a variable amount of uncertainty on the estimate of $q$, which we can be defined as a percentage of unity, as $p$ has been normalized to 1.

First, we use Monte Carlo simulations to simulate the error distribution for a given material, which we define as having a "true" $m_L/m_S = 0.05$, corresponding to the case where $q = 0.8p$. We then assume that the experimental errors in estimating $q$ can be modeled as arising from a normal distribution having an expected value of 0.05 and a standard deviation $\sigma_q$ that varies depending on the noise level of the experiment. Again, $\sigma_q$ is defined as a percentage of unity for simplicity. Here, we simulate cases where $\sigma_q$ ranges from 5% to 40%. For each $\sigma_q$, we draw 1000 random samples ($q_i$) from this distribution given the corresponding $\sigma_q$. Each $q_i$ is converted into an observed $m_L/m_S$ using **Equation (2)** and the results for all $\sigma_q$ are summarized in **Fig 5**.

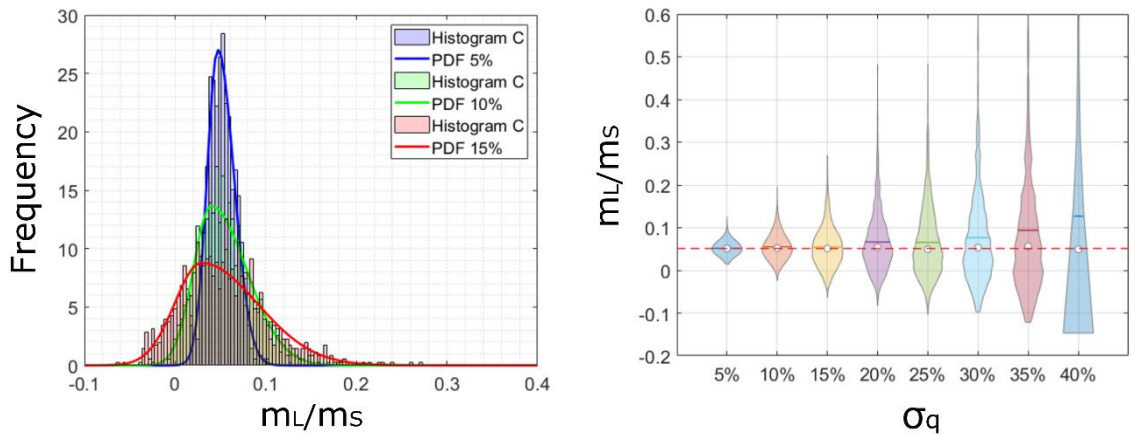

**Fig 5**. Left: Histograms with fitted distribution functions showing the $m_L/m_S$ values for the same material (here called 'C') with different noise characteristics, right: Violin plot showing the distribution of $m_L/m_S$ for various noise levels.

In **Fig 5 (a)**, we observe that, despite assuming a normal distribution of errors on $q$, the retrieved distribution of $m_L/m_S$ values is asymmetrical and strongly skewed to higher values. While this is not very pronounced for lower $\sigma_q$, it becomes increasingly significant as the estimate of $q$ becomes more uncertain. As described in the experimental results, a normal distribution function does not adequately model this empirical distribution, and its use will result in an estimate of $m_L/m_S$ that is biased towards larger values for higher $\sigma_q$. We therefore use an epsilon skew normal distribution (ESD) introduced above. The three ESD PDF models for these data are provided as solid lines in **Fig 5 (a)**. To visualize the influence of the bias in these data, we use a violin plot, presented in **Fig 5 (b)**. The colored bars show the mean values whereas the dots show the median of each dataset. Like the experimental observations, a simple median predicts the true $m_L/m_S$ for this material much better than the mean, and that this is most pronounced for the case when the uncertainty in $q$ is highest.

We now consider the case where the experimental uncertainty is constant but three different materials are measured, with $m_L/m_S = 0.38$, $0.17$, and $0.05$ for materials A, B, and C, respectively. For each material, we assume $\sigma_q = 0.10$ and use the same Monte Carlo approach described above to estimate the material-specific error distribution.

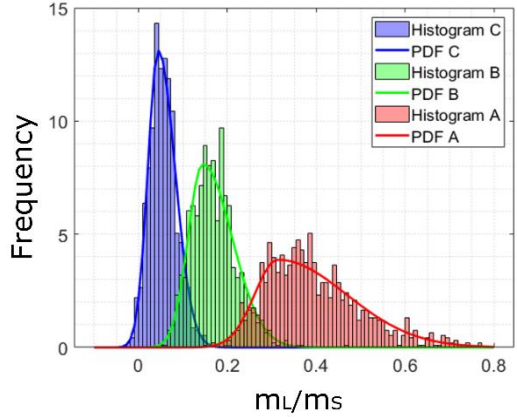

**Fig 6.** Histograms fitted with distribution functions showing the $m_L/m_S$ values for three different materials, considering the same noise characteristics.

The results of this analysis are presented in **Fig 6**. When the true value of $m_L/m_S = 0.05$, we observe that both the ESD skew and scale are relatively low. However, for higher $m_L/m_S$, both of these parameters increase markedly. This finding underscores the need for very high SNR experiments when the true value of $m_L/m_S$ is unknown but could potentially be somewhat large, as such materials will inherently carry a higher uncertainty in the estimation of their $m_L/m_S$ values.

## 4. Discussion

An EMCD signal is characterized by inverse intensity differences observed at $L_3$ and $L_2$ energy loss edges in the EELS spectra of transition metals. The relatively blurred profile and inherently lower signal at $L_2$ energy loss edge makes it more challenging to experimentally retrieve the $L_2$ EMCD signal. This is evident from many experimental studies where the $L_2$ EMCD signal is either very weak or just disappears under the noise level [4] [26] [27] [48] [49]. While this is appropriate for a qualitative study, the quantitative analysis requires the detection of a clear signal at both $L_3$ and $L_2$ edges with sufficient SNR. As confirmed by both the experimental results and simulations shown above, for noisy EMCD signals which is the case in most of the experiments, the errors in quantification are not symmetric while considering the mean as the reported value and they are strongly skewed towards larger values, suggesting asymmetric error bars. It is important to note that this asymmetry in the error bars is mainly dependent on the $SNRL_2$ which means a higher SNR at $L_3$ edge does not guarantee a reliable quantitative result. So, we suggest taking $SNRL_2$ as the criterion to determine the reliability of quantification. We have plotted the skewness in $m_L/m_S$ distributions obtained from the experimental data as a function of $SNRL_2$ in **Fig 7**. For $SNRL_2 > 5$ threshold, the error bars are symmetric, and the results follow a Normal distribution. This threshold is somehow similar as defined by rose criterion [50] which states that a signal must be 5 standard deviations above the background for a reliable detection. In practice, it is challenging to achieve this SNR, particularly at $L_2$ edge. If the experiment is done in TEM mode where one or few spectra are acquired, there is a fair possibility to get an over-estimated $m_L/m_S$ value as a result of quantification unless the signal fulfils the criterion defined above. A better approach is to take multiple samples as done in a STEM-EMCD experiment. In this way, not only the electron dose can be efficiently distributed over the region of analysis, but a high SNR can be obtained by integrating multiple spectra in the dataset. If the quantitative process is run on individual spectra, the median of multiple measurements should be considered as the reference value as it gives a close estimate to the true $m_L/m_S$ value. Moreover, the q-slope in **Fig 4** can be very non-linear due to multiple effects such as correlated noise and fixed pattern noise, producing further uncertainties in quantification. We suggest using gain averaging [51] during acquisition to remove these artefacts.

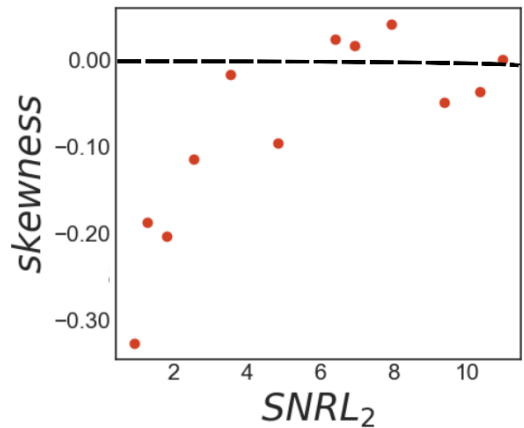

**Fig 7**. Skewness in the distribution of $m_L/m_S$ values plotted against SNR of $L_2$ for the experimental data.

## 5. Conclusions

We have employed bootstrapping to estimate errors in quantitative STEM-EMCD experiments. We report that the resulting $m_L/m_S$ values are biased towards larger values for noisy signals. This bias is not only dependent on noise but also material dependent and the materials having larger intrinsic $m_L/m_S$ values show higher bias for the same noise levels. The skewness in the distribution of resulting $m_L/m_S$ values suggest using an asymmetric error bar for noisy signals whereas for multiple measurements, a median closely represents the true value.

## ACKNOWLEDGEMENTS

H.A. acknowledges funding from Swedish Research Council (project nr. 2021-06748) for financial support. H.A, C-W.T and T.T acknowledge the financial support from Swedish Foundation for Strategic Research SSF (ITM17-0301). T.T acknowledges Swedish Research Council (project nr. 2016-05113). J.R. acknowledges Swedish Research Council (project no.2021-03848), Carl Tryggers Foundation and STINT for financial support.